\documentclass[preprint]{aastex631}

\usepackage{amsmath,mathrsfs,bm}
\usepackage{graphicx,xcolor}
\usepackage{natbib}

\DeclareMathAlphabet{\tens}{OT1}{cmss}{bx}{n}

\begin{document}


\title{Image instabilities and polarization cross-talk}
\author{R.~Casini \& A.~G.~de Wijn}
\affil{HAO, NCAR, P.O. Box 3000,
Boulder, 80307-3000, CO, U.S.A.}

\begin{abstract}
We expand on our previous study of the impact of atmospheric seeing 
on polarization cross-talk, and show how the formalism that was developed in that work
can be applied to treat the case of spatial modulators of polarization. 
Beside formally demonstrating how the problem of cross-talk is 
fully eliminated in such devices, we also gain insight on the 
meaning of polarimetric noise of temporal modulation schemes in 
the limit of very high modulation frequency. 
We also describe the problem of spectrograph instabilities, and how 
the spectral gradients that are naturally associated with a line 
spectrum feed into the problem of polarimetric errors induced by 
mechanical vibrations, thermal drifts, and pointing jitter.
Finally, we show how this formalism can be used to estimate 
the contribution of polarization cross-talk to the errors on the
elements of the 4$\times$4 Stokes response matrix, for the 
purpose of producing realistic error budgets for polarimetric 
instrumentation.
\end{abstract}


\section{Introduction} \label{sec:intro}

One of the main challenges in achieving high-precision polarimetry 
in the remote sensing of astrophysical targets is the mitigation of 
spurious polarization created by image instabilities, whether these 
are caused by the natural time evolution of the target, 
disturbances in the Earth atmosphere (\emph{seeing}) affecting the 
signals acquired by ground-based instrumentation, or the pointing 
jitter of spacecrafts. The most common implementations of polarimetry 
rely on \emph{temporal} modulation schemes of the incoming radiation, 
which employ birefringent optics cycling through a specified set of 
configurations (the \emph{polarization modulation cycle}) 
to encode the polarization of the target into a sequence of different 
intensity signals. Depending on the frequency of this modulation cycle 
and the characteristic frequency distribution of the image instabilities 
(e.g., the power spectral density, or PSD, of spacecraft jitter or of 
atmospheric turbulence), time correlations between the modulation 
cycle and such instabilities may be realized, biasing the inference of 
the true state of polarization of the target upon demodulation of the 
acquired signals.

In \citeauthor{Ca12} (\citeyear{Ca12}, hereafter, Paper~I; see also
Erratum), we
explicitly considered the problem of image instabilities due to 
atmospheric seeing, described as a stationary random process, and 
developed a general formalism for estimating the induced errors 
on the Stokes vector measurements of an observed target, for typical
implementations of temporal modulation schemes (specifically, 
for either single- or dual-beam polarimeters, with either stepped or 
continuously rotating polarization modulators).

%
In the present work, we show how to extend the formalism of Paper~I to 
model the effects of image instabilities in polarimetric instruments 
that adopt \emph{spatial modulators}, i.e., 
polarimetric devices where all signals needed to realize a complete 
measurement of the polarization state of the target are 
simultaneously acquired (whether on different detectors, or different
areas of the same detector). Quite naturally, the polarization errors 
associated with such devices are shown to correspond to the asymptotic 
lower bound of the cross-talk errors produced in temporal modulation 
schemes, in the limit of very high modulation frequency.

Next, we derive formulas that allow us to estimate the polarization
errors induced by geometric instabilities of slit-based spectrographs. 
These errors are commonly caused by mechanical vibrations and thermal
drifts in the instrument, and are driven by the presence of signal gradients 
at the detector, which naturally manifest themselves when spectroscopically 
observing a target that emits a line spectrum. In particular, such errors 
can be produced even when the target has a spatially uniform brightness 
(in the case of extended objects), or when the telescope pointing is
sufficiently precise to stably maintain the position of the target at 
the entrance slit of the spectrograph.

Finally, we show how the various noise contributions to Stokes measurements 
estimated with this formalism can practically be used to inform the polarimetric 
error budget of a spectro-polarimetric instrument, which is a necessary
and critical step of any instrument design, whether for ground-based or
space-borne observations. 


\section{A short review of the seeing-induced cross-talk formalism} \label{sec:summary}

In Paper~I we considered the case of an extended target emitting a radiation 
field with constant polarization properties, described by the Stokes parameters 
$S_i$ ($i=1,2,3,4$, respectively, for $S_i=I,Q,U,V$), but allowing for the 
possibility of spatial gradients $\bm{\nabla}S_i$ over the extension of the 
target. In the presence of image instabilities that can be modeled as a 
\emph{stationary random process}, with zero average, and described by a power
spectral density (PSD) $S(\omega)$, where $\omega$ is the angular
frequency spanning the domain of the random process, we had derived 
formal expressions for the expectation value and variance of the 
\emph{observed} Stokes parameter $S_i'$ (cf.\ Eqs.~(14) and (40) 
of Paper~I), respectively,
\begin{eqnarray} 
\label{eq:mean}
\bar{S_i'}&=&\kappa\Delta t\,S_i\;, \\
\label{eq:variance}
\sigma^2(S_i')
&=&\frac{\kappa^2\Delta t^2}{2\pi}
	\int_{-\infty}^{+\infty} S(\omega)\,
	\bigl(
	\tens{\tilde H}(\omega)\tens{G}\tens{\tilde H}^\dagger(\omega)
	\bigr)_{ii}\;\textrm{d}\omega\;,
\end{eqnarray}
where $\Delta t$ is the exposure time for each modulation state, and $\kappa$ some dimensional constant. In Eq.~(\ref{eq:variance}) we indicated with $\tens{A}^\dagger$ the operation of Hermitian conjugation of a matrix $\tens{A}$, i.e., $\tens{A}^\dagger=(\tens{A}^t)^\ast=(\tens{A}^\ast)^t$, where $\tens{A}^t$ is matrix transposition, and $\tens{A}^\ast$ is complex conjugation.

Under the assumption that the image motions in the two orthogonal directions of the image plane are independent random processes described by the same PSD (see discussion following Eq.~(24) of Paper~I), the 4$\times$4 matrix $\tens{G}$ is given by the product of the 2$\times$4 Jacobian matrix $\bm{\nabla S}$ with its transpose, i.e., $(\tens{G})_{ij}=\bm{\nabla}S_i\cdot\bm{\nabla}S_j
=\partial_k S_i\,\partial_k S_j$ ($k=1,2; i,j=1,\ldots,4$), where the Stokes gradients are determined by the spatial distribution of the radiation source measured at the detector. 
Therefore, by construction, $\tens{G}$ is symmetric, non-negative, and of rank $\le
2$ (i.e., it has at least two null eigenvalues, while any 
non-null eigenvalue is positive).\footnote{If the
gradient of the Stokes vector in one of the two dimensions of the
detector is zero, then the rank of $\tens{G}$ drops to 1.}

The 4$\times$4 matrix $\tens{\tilde H}(\omega)$ 
consists of Fourier transforms 
of the polarization-modulation time sequence, and is dimensionless. 
This last property follows from the definition of the originating matrix 
$H_{ij}(\xi)$ (see Eq.~(39) of Paper~I), whose units are inverse time, 
and from Parseval's theorem. 
Therefore, the normalized gradients have the units of $\Gamma(0)^{-1/2}$ 
(e.g., arcsec$^{-1}$), where $\Gamma(t)$ is the autocorrelation function
of the image motion (see Eq.~(33) of Paper~I). 


We note that the integral over the frequency range 
$(-\infty,+\infty)$ 
in Eq.~(\ref{eq:variance}) can be expressed more conveniently as an equivalent integral restricted to just the positive frequency range $[0,+\infty)$, in virtue of the parity of the integrand. This property was already pointed out without formal demonstration in Paper~I (see discussion after Eq.~(40)). Here we provide such a demonstration in the Appendix.
Using this property, and introducing the intensity normalized gradient matrix $\tens{g}\equiv\tens{G}/S_1^2$, from Eqs.~(\ref{eq:mean}) and (\ref{eq:variance}) we find
\begin{equation} \label{eq:error}
\biggl[\frac{\sigma(S_i')}{\bar S_1'}\biggr]^2
=\frac{1}{\pi}
	\int_{0}^{\infty} S(\omega)\,
	\bigl(
	\tens{\tilde H}(\omega)\tens{g}\tens{\tilde H}^\dagger(\omega)
	\bigr)_{ii}\;\textrm{d}\omega\;.
\end{equation}

For a stepped modulator with $n$ states, it is straightforward to cast 
Eq.~(42) of Paper~I into matrix form. For this purpose, we introduce 
the diagonal phase matrix of rank $n$,
\begin{equation} \label{eq:phasemat}
\Phi_{ij}(\omega)=\delta_{ij}\;\mathrm{e}^{-\mathrm{i}(j-1)\omega
T/Nn}\;,
\end{equation}
where $N$ is the number of modulation cycles in the polarimetric
sequence, and $T$ is the total duration of the sequence 
(dwell time). Using this phase matrix
we can write
\begin{equation} \label{eq:Hmat}
\tens{\tilde H}(\omega)
=\textrm{sinc}(\omega\Delta t/2)
	\frac{\textrm{sinc}(\omega T/2)}
	{\textrm{sinc}(\omega T/2N)}\,
\tens{D}\tens{\Phi}(\omega)\tens{M}\;,
\end{equation}
so the matrix element in the integrand of Eq.~(\ref{eq:error}) becomes
\begin{equation} \label{eq:integrand}
\bigl(
\tens{\tilde H}(\omega)\tens{g}\tens{\tilde H}^\dagger(\omega)
\bigr)_{ii}
=\textrm{sinc}^2(\omega\Delta t/2)
	\frac{\textrm{sinc}^2(\omega T/2)}
	{\textrm{sinc}^2(\omega T/2N)}\,
\bigl(
\tens{D}\tens{\Phi}(\omega)\tens{M} \tens{g}
\tens{M}^t\tens{\Phi^\ast}(\omega)\tens{D}^t
\bigr)_{ii}\;.
\end{equation}

Because $S(\omega)$ is non-negative, the positivity of the cross-talk variance, Eq.~(\ref{eq:error}), implies that the matrix elements of Eq.~(\ref{eq:integrand}) must also be non-negative everywhere. This property is also formally demonstrated in the Appendix.

With this summary of the  seeing-induced 
cross-talk formalism presented in Paper~I, and the fundamental properties of Eq.~(\ref{eq:error}) being 
demonstrated, we are now in a position to apply the formalism to the study of spatial modulators.

\section{Spatial modulators} \label{sec:spatial}

A spatial modulator is a polarimetric device where all  
modulation states required to complete a polarization measurement of a target are acquired 
simultaneously, albeit they need to be imaged on different detectors 
or different areas of the same detector.
In this section we want to show how the estimation Eq.~(\ref{eq:error}) of polarimetric cross-talk can be extended to this type of device.

For the sake of simplicity, we assume that the different modulated images can be 
perfectly co-registered, and that a properly accurate flat-field of the 
detector(s) can be ensured.
We also assume that the polarized radiation emitted by the target does
not evolve over the duration of the signal-integration sequence.

We then follow the same derivation as in Paper~I, Sect.~6, to estimate the 
polarimetric errors produced over a series of $N$ repeated measurements. 
In particular, Eqs.~(39) and (40) of Paper~I still apply with the following 
straightforward modifications, which are implied by the nature of the 
spatial modulation process:
\begin{eqnarray*}
	&m_j(\xi+t_k)\equiv m_{kj}\;, &\qquad \hbox{(modulation independent of $t$)} \\
	&t_k=0\;. &\qquad \hbox{(instantaneous modulation ``cycle'')}
\end{eqnarray*}
Equations~(39) and (42) of Paper~I then become
\begin{eqnarray}
	\label{eq:Hij}
H_{ij}(\xi)&=&\frac{1}{\Delta t}\,\delta_{ij} \sum_{l=0}^{N-1}
	\delta(\xi-l\Delta t/r)\ast\Pi_{\Delta t}(\xi)\;, \\
	\label{eq:FTHij}
\tilde H_{ij}(\omega)&=&\delta_{ij}\;
	\textrm{sinc}(\omega\Delta t/2)\,
	\frac{\textrm{sinc}(\omega T/2)\,}
	{\textrm{sinc}(\omega T/2N)}\;,
\end{eqnarray}
where $\Delta t$ and $T$ are again the camera exposure time and the 
total duration of the polarimetric sequence, respectively, but now 
$N$ represents the total number of exposures, i.e., $T=N\Delta
t/r$, where $0<r\le 1$ is the camera duty cycle.

\begin{figure}[t!]
\centering
\includegraphics[scale=.57]{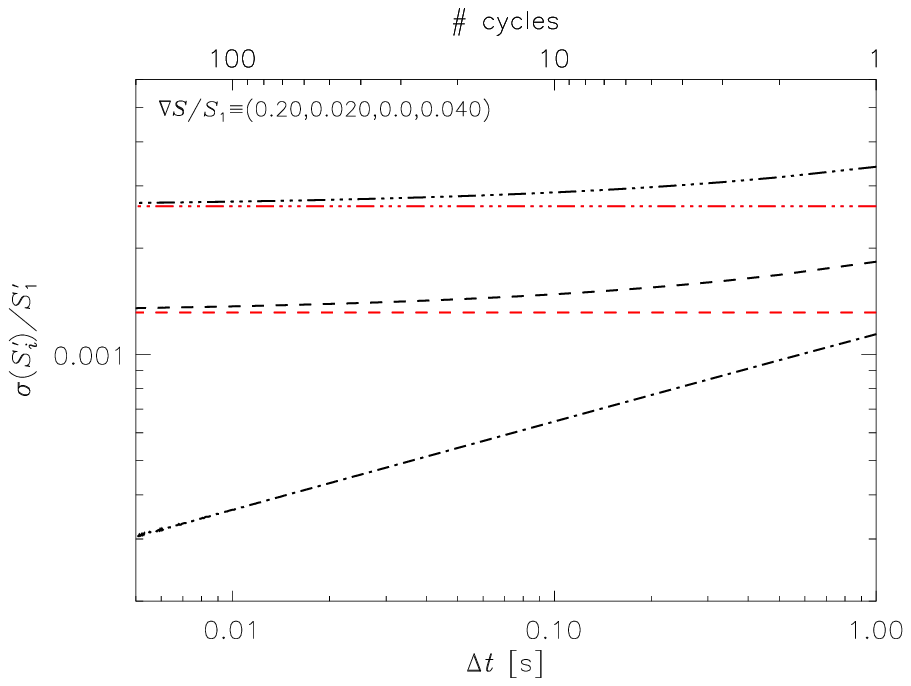}
\includegraphics[scale=.57]{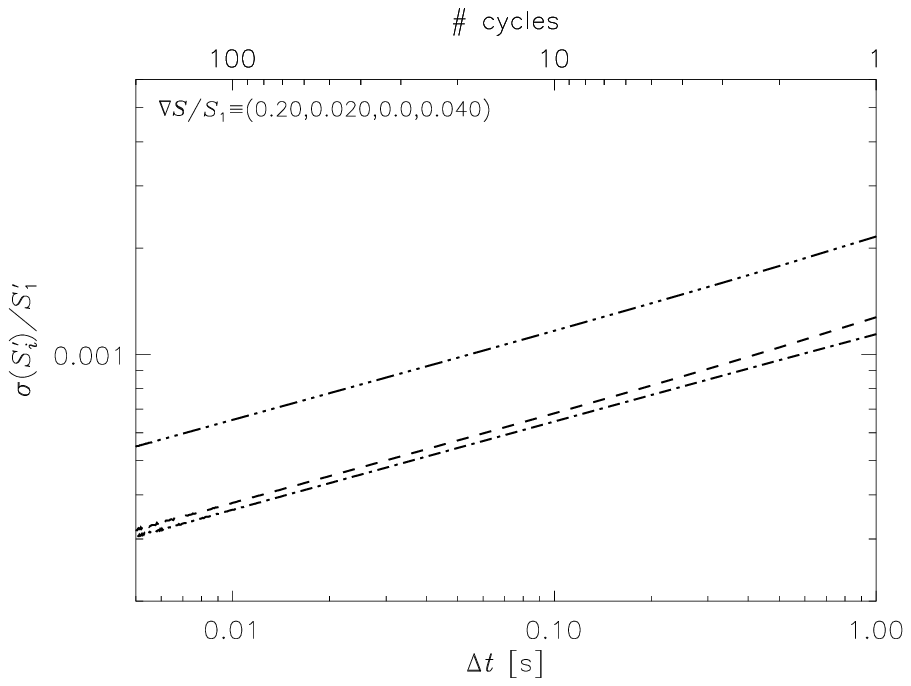}
\caption{\label{fig:jitter}
Polarization modulation errors induced by a ``jitter'' power spectrum $1/\nu^{1/2}$, 
in the presence of Stokes vector gradients in the incoming signal. In this example, the PSD of the jitter motion, $S(\nu)$, ranges between 0.01\,Hz and 1\,kHz, with a rms amplitude of 1\,arcsec; the polarization modulator is a simple linear retarder with $150^\circ$ of retardance, continuously rotating over the modulation cycle consisting of 8 states; the integration time is $T=8$\,s, with a camera duty cycle of 100\%; finally, we assumed the optimal demodulation of the signals in a dual-beam polarimeter design. 
The normalized Stokes gradient 
vector is indicated in the plots, and is expressed in units of arcsec$^{-1}$. It describes a worst-case scenario for solar polarimetric observations, with an intensity contrast of 20\% over the solar-granulation length scale, and assuming 10\% linear polarization and 20\% circular polarization. In this example, the linear polarization is assumed to be completely folded into Stokes $Q$, while Stokes $U$ is set to zero. 
\emph{Left:} total modulation polarimetric errors; \emph{Right:} polarization 
cross-talk errors only. 
The different curve styles indicate different Stokes parameters: $Q$ (dash), $U$ (dot-dash), $V$ (dash-triple-dot). 
The red curves in the left panel show the limit case of a spatial 
modulator (i.e., all 8 modulation states are 
acquired simultaneously; see Eq.~(\ref{eq:spat_error})). We note the presence in both panels of Stokes-$U$ polarization cross-talk  despite having assumed Stokes $U$ and its gradient in the signal to be zero. This explains the absence of a corresponding diagonal (spatial modulator) contribution in the left panel, and the fact that the Stokes-$U$ errors are exactly the same in the two panels, and completely due to polarization cross-talk from the non-zero Stokes parameters $Q$ and $V$. } 
\end{figure}

Thus, for the polarimetric error produced in a spatial modulator, 
Eq.~(\ref{eq:error}) simply gives
\begin{equation} \label{eq:spat_error}
\biggl[\frac{\sigma(S_i')}{\bar S_1'}\biggr]^2_\mathrm{sp.mod.}
	=\frac{\tens{g}_{ii}}{\pi}
	\int_{0}^{\infty} S(\omega)\,\tilde H_{ii}^2(\omega)\;
	\textrm{d}\omega\;,
\end{equation}
with $\tens{g}_{ii}=|\bm{\nabla}S_i/S_1|^2$ and $\tilde H_{ij}(\omega)$ given by 
Eq.~(\ref{eq:FTHij}). 
This shows that the polarimetric error on the Stokes parameter $S_i$ is 
exclusively determined by the PSD of the image motion, weighed by the 
corresponding gradient $\bm{\nabla}S_i$ of the target. Therefore,
a spatial modulator does \emph{not} introduce polarization 
cross-talk, since the gradient matrix $\tens{g}$ only enters those 
expressions through its diagonal elements, and is also fully decoupled 
from the Fourier components $\tilde H_{ij}(\omega)$ of the modulation 
scheme.

We can interpret these ``diagonal'' polarimetric errors simply in terms 
of the loss of spatial information caused by the image motion during the 
acquisition of the polarization signals, i.e., the spatial smearing of 
the Stokes vector of the target caused by integrating the signals over the
random realizations of the image motion. Thus, uncorrected image instabilities 
in a spatial modulation scheme will result in each of the Stokes 
parameters of the source being spatially averaged over the 
characteristic length scale of the image motion, which is estimated
by its rms amplitude. On the other hand, no 
cross-talk between different Stokes parameters is produced in this case, 
because of the simultaneity of the full polarimetric information 
acquired with every camera frame.

Figure~\ref{fig:jitter} demonstrates these conclusions by numerically
comparing the polarization errors of temporal and spatial modulation
schemes in the presence of image jitter with a $1/\nu^{1/2}$ power spectrum, 
assuming a camera duty cycle $r$ of 100\%. In this case, Eq.~(\ref{eq:FTHij}) 
shows that the polarimetric errors on the Stokes measurements
using a spatial modulator are only a function of the total integration 
time $T$, but not of the number of exposures $N$ (left panel). However, a small 
dependence on $N$ is introduced for $r<1$, with smaller errors being
produced for larger values of $N$ as expected. In contrast, 
the cross-talk errors produced by a temporal modulation scheme 
decrease with the modulation frequency (essentially like $1/\sqrt N$, when the image jitter can be modeled as a stationary random process, as it is assumed in this work), while they are identically zero in the case of a spatial modulator (right panel). Because of this, the total polarimetric errors in the left plot decrease as a function of $N$ for a temporal
modulation scheme, converging to the values corresponding to the case of
a spatial modulator, which are attained in the limit of very large modulation frequency 
(very large $N$, for a given value of $T$). These non-zero limit values 
correspond to the unavoidable loss of 
spatial resolution of the Stokes parameters of the target, caused by 
the averaging of the randomly displaced target, with an rms 
amplitude determined by the autocorrelation function of the image motion.

To conclude this analysis, it is important to remark that, under the 
assumption that the image motion is a stationary random process with 
zero average (i.e., the true mean position of the target does not drift 
over time), increasing the length of the polarimetric sequence (i.e., 
the total integration time) reduces the width of the error distribution 
around the true position, as a result of the improved statistics of 
the random realizations of the image motion. For this reason, extending
the integration time of Stokes measurements has the overall effect of 
reducing the polarimetric errors in both temporal (for a fixed exposure 
time) and spatial modulation schemes.

\newpage

\section{Spectral effects} \label{sec:spectral}
In the case of spectrally resolved polarimetric observations that employ
traditional slit-based spectrographs, one must take into
account the possible mechanical and thermal instabilities of the 
spectrograph, and how these may induce polarization cross-talk  
over the widely different time scales of such instabilities, because 
of the presence of \emph{spectral} gradients 
of the signals, which are of course non-vanishing in any observation
of spectral lines. These instabilities typically range from the 
${\sim}10^{-1}{-}10^2$\,Hz of mechanical vibrations, down to the 
${\sim}10^{-4}$-$10^{-2}$\,Hz characteristic of pointing and thermal drifts.

If we indicate with $\partial_\lambda$ the spectral derivative
operator, we evidently have 
$\partial_x=(\partial\lambda/\partial x)\,
\partial_{\lambda}$, where $x$ here represents the spatial displacement 
along the direction of spectral dispersion. 
For a grating-based spectrograph, spectral displacements at
the detector can be caused by changes in both the incidence and
diffraction angles, $\alpha$ and $\beta$, respectively. 
Using the grating equation,
\begin{displaymath}
\sin\alpha+\sin\beta=m\lambda/d\;,
\end{displaymath}
where $m$ and $d$ are the grating order and groove density,
respectively, and $\lambda$ is the observed wavelength, we have
\begin{eqnarray} \label{eq:gratingformula}
\frac{\mathrm{d}\lambda}{\lambda}
&=&\frac{\cos\alpha\,\mathrm{d}\alpha+\cos\beta\,\mathrm{d}\beta}{\sin\alpha+\sin\beta}
\nonumber \\
&=&\frac{\cos\alpha\,\mathrm{d}x_\alpha/f_{\rm coll}
+\cos\beta\,\mathrm{d}x_\beta/f_{\rm cam}}{\sin\alpha+\sin\beta}\;,
\end{eqnarray}
where $f_{\rm coll}$ and $f_{\rm cam}$ are, respectively, the 
focal lengths of the collimator and camera systems of the 
spectrograph. Therefore,
\begin{eqnarray}
\label{eq:del_alpha}
\partial_{x_\alpha}
&=&\frac{\partial\lambda}{\partial x_\alpha}\,\partial_\lambda
=\frac{\lambda}{f_{\rm coll}}\,\frac{\cos\alpha}{\sin\alpha+\sin\beta}\,
\partial_\lambda\;, \\
\label{eq:del_beta}
\partial_{x_\beta}
&=&\frac{\partial\lambda}{\partial x_\beta}\,\partial_\lambda
=\frac{\lambda}{f_{\rm cam}}\,\frac{\cos\beta}{\sin\alpha+\sin\beta}\,
\partial_\lambda\;.
\end{eqnarray}

For a given spectrograph configuration (fixed $\alpha$), if we indicate with $\delta_R$ the minimum spatial width that resolves the spectrograph 
profile at the detector (typically, this will require $\delta_R$ to span 
between 2 to 3 pixels of the detector, for a critically sampled
instrument profile, based on the Nyquist criterion), 
we can use Eq.~(\ref{eq:gratingformula}) to estimate the resolving power
$R$ of the spectrograph,
\begin{equation}
\frac{1}{R}\equiv\frac{\mathrm{d}\lambda}{\lambda}\biggr|_\alpha
=\frac{\cos\beta}{\sin\alpha+\sin\beta}\,\frac{\delta_R}{f_{\rm cam}}\;.
\end{equation}
This allows us to rewrite Eqs.~(\ref{eq:del_alpha}) and
(\ref{eq:del_beta}) explicitly in terms of the spectral resolution of
the instrument,
\begin{eqnarray}
\label{eq:del_alpha_R}
\partial_{x_\alpha}
&=&\frac{r_{\rm sp}\,r_{\rm ana}}{R}\,\frac{\lambda}{\delta_R}\,
\partial_\lambda\;, \\
\label{eq:del_beta_R}
\partial_{x_\beta}
&=&\frac{1}{R}\,\frac{\lambda}{\delta_R}\,
\partial_\lambda\;,
\end{eqnarray}
where in Eq.~(\ref{eq:del_alpha_R}) we introduced the spectrograph and
anamorphic magnifications, 
$r_{\rm sp}=f_{\rm cam}/f_{\rm coll}$ and 
$r_{\rm ana}=\cos\alpha/\cos\beta$, respectively.

Equations~(\ref{eq:del_alpha}) and (\ref{eq:del_beta}), or their equivalent forms (\ref{eq:del_alpha_R}) and (\ref{eq:del_beta_R}), allow us to
estimate the spatial gradients that must be used in Eq.~(\ref{eq:variance}) to evaluate the effects of spectrograph
instabilities on polarimetric cross-talk in the presence of spectral 
gradients naturally associated with the intensity distribution of the observed signal across a line spectrum.
The PSDs of the image jitter at the entrance slit and at the detector, 
which correlate with the spatial gradients $\partial_{x_\alpha}$ and
$\partial_{x_\beta}$, respectively, can be estimated via a
finite-element analysis of the instrument design, or measured through 
targeted environmental tests of a prototype instrument.

\begin{figure}[t!]
    \centering
    \includegraphics[width=.88\textwidth]{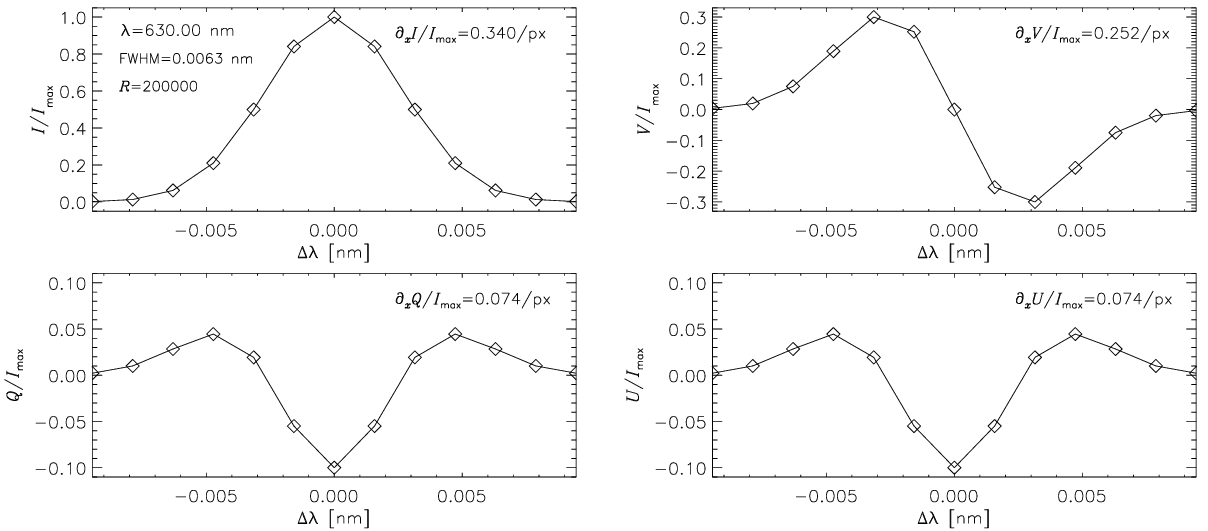}
    \includegraphics[width=.88\textwidth]{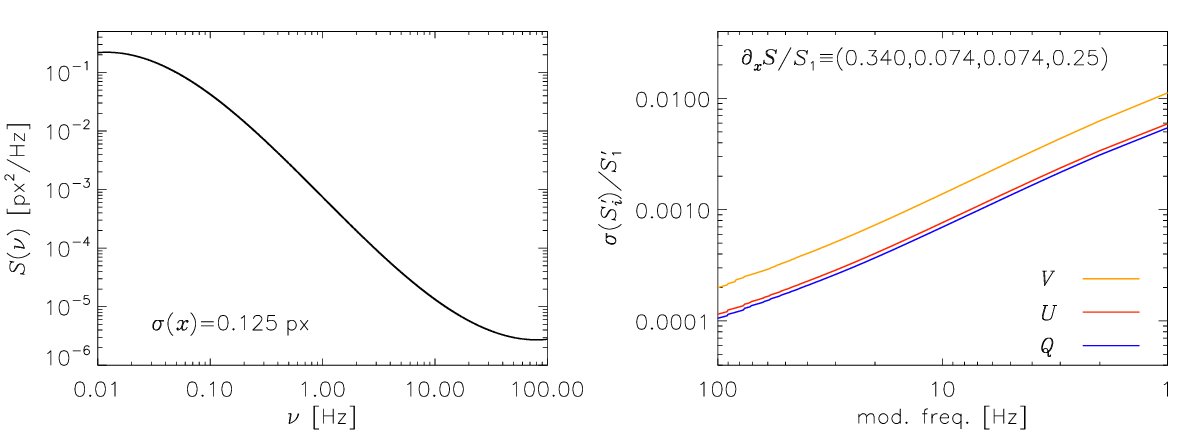}
    \caption{\emph{Top two rows:} Example of spectrally resolved Stokes profiles of a narrow line (FWHM $\sim 6.3$\,pm), observed with a resolving power of 200\,000 (2-px critical-sampling), and assuming maximum linear polarization of 10\% and circular polarization of 30\%. The spectral gradients reported inside the panels are obtained through Eq.~(\ref{eq:del_beta_R}), and must be used in Eq.~(\ref{eq:error}) to estimate the corresponding polarimetric cross-talk errors. 
    \emph{Bottom row: (left)} modeled PSD of the image jitter at the detector in the spectral direction, delivering a target 1/8\,px rms \citep{DW22}; \emph{(right)} polarization cross-talk errors induced by the same PSD, for a spectrograph configuration representing the observation of the synthetic Stokes profiles shown at the top, assuming a total integration time of 10\,s and a 10-state modulation cycle.)}
    \label{fig:ViSP}
\end{figure}

To demonstrate the effects of spectrograph vibrations, in Fig.~\ref{fig:ViSP} we modeled the polarimetric cross-talk induced by spectral jitter with a characteristic vibration PSD of a spectrograph. Measurements of this kind were taken during the site-acceptance testing of the Visible Spectro-Polarimeter \cite[ViSP;][]{DW22} at the NSF Daniel K. Inouye Solar Telescope \cite[DKIST;][]{Ri20}, by acquiring time series of the position of the image of the slit hairlines on the detector, using the spectrograph in 0th-order mode, i.e., with the grating acting as a non-dispersing mirror. The PSD used in the example of Fig.~\ref{fig:ViSP} (bottom-left panel) closely reproduces the typical shape of such observed vibration power spectra, and it was appropriately scaled so as to produce a spectral jitter of 1/8\,px rms, as prescribed by the instrument requirements of the ViSP \citep{DW22}.

The first two rows of Fig.~\ref{fig:ViSP} simulate the Stokes profiles of a narrowly resolved spectral line observed with a typical resolving power for the ViSP of $R=200\,000$, along with an estimation of the normalized spectral gradients in units of px$^{-1}$, assuming a maximum of 10\% linear polarization and 30\% circular polarization. 
The estimated values of the spectral gradients for such a representative observation are used in Eq.~(\ref{eq:error}) to compute the polarimetric cross-talk produced by a modulation cycle with 10 states for different values of the modulation frequency, over a total integration time of 10\,s. The resulting polarization cross-talk errors are shown in the bottom-right panel.



Equations~(\ref{eq:del_alpha}) or (\ref{eq:del_alpha_R}) can also be used 
to determine the impact of pointing jitter of a telescope feeding the 
spectrograph, by observing that
\begin{equation} \label{eq:pointing}
\partial_\varphi
=\frac{f_{\rm tel}}{f_{\rm coll}}\,
	\frac{\lambda \cos\alpha}{\sin\alpha+\sin\beta}\,\partial_\lambda
=f_{\rm tel}\,\frac{r_{\rm sp}\,r_{\rm ana}}{R}\,\frac{\lambda}{\delta_R}\,
\partial_\lambda\;,
\end{equation}
where $f_{\rm tel}$ is the focal length of the imaging telescope, and 
the pointing angle $\varphi$ is expressed in radians. For a point
source, 
it is appropriate to further scale the gradient of Eq.~(\ref{eq:pointing}) 
by the Strehl ratio of the telescope, whereas in the observation of an 
extended source, the appropriate scaling factor becomes the intensity 
contrast of the target.

\section{Polarization cross-talk terms as Stokes response matrix errors}
\label{sec:error}

We note that the formalism of Paper~I was developed assuming 
that the demodulation matrix $\tens{D}$ is given by the Moore-Penrose 
inverse of the modulation matrix $\tens{M}$, 
i.e., $\tens{D}=(\tens{M}^t\tens{M})^{-1}\tens{M}^t$, which is known to 
provide the optimal demodulation of the signals acquired through a given
modulation scheme with matrix $\tens{M}$ \citep{dTC00}. For computational 
reasons (e.g., fast data processing onboard space missions prior to 
downlink), it is often convenient to employ a simplified demodulation 
scheme $\tens{D}_0$ that produces a ``provisional'' Stokes vector 
$\bm{S}_0$ from the observed modulated signals, which must then be 
post-processed to obtain the true Stokes vector $\bm{S}$. The only 
necessary condition is that the 4$\times$4 matrix $\tens{D}_0\tens{M}$ 
be invertible.  In fact, in such a case,
\begin{equation}
\bm{S}_0=\tens{D}_0(\tens{M}\,\bm{S})
	\quad\to\quad
\bm{S}=(\tens{D}_0\tens{M})^{-1}\,\bm{S}_0\;.
\end{equation}
It is customary to introduce the \emph{polarimeter response matrix} $\tens{X}\equiv\tens{D}_0\tens{M}$ \cite[e.g.,][]{Ic08}, through which we can define an alternative, 
non-optimal, demodulation matrix $\tens{\tilde D}\equiv\tens{X}^{-1}\tens{D}_0$. Because
$\tens{\tilde D}\tens{M}=\tens{I}$, the whole formalism developed in Paper~I is 
directly applicable also to the case of non-optimal demodulation schemes. Obviously, $\tens{X}$ becomes the identity matrix in the optimal-demodulation case.

The concept of polarimeter response matrix is also useful for creating the 
polarimetric error budget of an instrument. To see this, we rewrite 
Eq.~(\ref{eq:mean}) in matrix form
\begin{equation} \label{eq:meanX}
\bm{S'}=\kappa\Delta t\,\tens{I}\bm{S}\;,
\end{equation}
where $\tens{I}$ is the 4$\times$4 identity matrix. If the actual 
modulation matrix $\tens{\hat M}$ at the time of the observation is 
different from the one determined via instrument calibration, 
$\tens{\hat M}\ne\tens{M}$, then we can write
\begin{equation}
	\bm{S'}=\kappa\Delta t\,(\tens{I}+\bm{\epsilon})\bm{S}\;.
\end{equation}
where $\bm{\epsilon}\equiv\tens{D}(\tens{\hat M}-\tens{M})$ is the
4$\times$4 error matrix, and $\tens{DM}=\tens{I}$.
If the entries in the error matrix are uncorrelated, by recalling
Eq.~(15) of Paper~I, 
the normalized variance on the observed Stokes parameter $S_i'$ due 
to the error matrix is
\begin{eqnarray} \label{eq:Xvar}
\biggl[\frac{\sigma(S_i')}{\bar S_1'}\biggr]^2
	&=&\frac{1}{S_1^2}\sum_{j,k=1}^4 E(\epsilon_{ij}S_j\,\epsilon_{ik}S_k)
	=\frac{1}{S_1^2}\sum_{j,k=1}^4 E(\epsilon_{ij} \epsilon_{ik})\,S_j S_k 
	=\frac{1}{S_1^2}\sum_{j=1}^4 E(\epsilon^2_{ij})\,S^2_j \nonumber \\
	&\equiv&\frac{1}{S_1^2}\sum_{j=1}^4 \epsilon^2_{ij}\,S^2_j\;,
\end{eqnarray}
where $E(x)$ is the expectation value of $x$.

The maximum acceptable error matrix $\bm{\epsilon}$ is typically determined by the 
desired polarimetric accuracy and the largest expected fractional polarization 
$|S_j|/S_1$, $j=2,3,4$, of the observations. As an example, the DKIST Error Budget 
definition prescribes the following error matrix
\begin{equation} \label{eq:errmat}
	\arraycolsep .25em
\bm{\epsilon}=\begin{pmatrix}
	10^{-2} &10^{-2} &10^{-2} &10^{-2} \\
	5{\times}10^{-4} &10^{-2} &5{\times}10^{-3} &5{\times}10^{-3} \\
	5{\times}10^{-4} &5{\times}10^{-3} &10^{-2} &5{\times}10^{-3} \\
	5{\times}10^{-4} &5{\times}10^{-3} &5{\times}10^{-3} &10^{-2}
	     \end{pmatrix}
\end{equation}
to specify the required accuracy of its
polarimetric instrumentation  \citep{Ha23}.  In a similar fashion, 
\cite{Ic08} specified the accuracy constraints for the SOT/SP 
instrument onboard the Hinode mission \citep{Ts08}.
Then, Eq.~(\ref{eq:Xvar}) allows us to set an upper limit to the acceptable 
accuracy level on the measurement of Stokes parameters. 

The
$\bm{\epsilon}$ matrix is used to establish a polarimetric error budget of a spectro-polarimeter, accounting for the 
contributions to polarimetric errors from various sources: modulation thermal instabilities; camera-modulator synchronization errors; image instabilities at various places along the optical train, such as image jitter at the entrance slit of a spectrograph, and mechanical instabilities of the spectrograph itself, such as grating jitter and spectrograph bench flexing (``seeing'' induced cross-talk; SICT). Under the condition that all these sources are independent, we may assume that the total error satisfies an inequality of the form
\begin{equation}
\biggl[\frac{\sigma(S_i')}{\bar S_1'}\biggr]^2 \ge
	\biggl[\frac{\sigma(S_i')}{\bar S_1'}\biggr]^2_\mathrm{mod}+
	\biggl[\frac{\sigma(S_i')}{\bar S_1'}\biggr]^2_\mathrm{sync}+
	\biggl[\frac{\sigma(S_i')}{\bar S_1'}\biggr]^2_\mathrm{SICT}+ \cdots
\end{equation}

We can use the formalism of these notes to derive an 
approximate expression for the contribution of image instabilities 
(SICT) to the matrix element $\epsilon_{ij}$. 
Recalling Eq.~(\ref{eq:error}) (cf.\ also Eq.~(40) of Paper~I), 
and neglecting the corrections introduced by the off-diagonal 
elements of the normalized gradient tensor $\tens{g}$,\footnote{The
contribution to polarization cross-talk associated with the 
off-diagonal elements of the gradient tensor was investigated 
in Paper~I. For the current purpose of providing a ROM estimation 
of the magnitude of the elements of the error matrix $\bm{\epsilon}$, 
it is sufficient to consider the predominant contribution from 
the diagonal elements of $\tens{g}$.}
we can write (see Eq.~(\ref{eq:Xvar})),
\begin{eqnarray*}
\biggl[\frac{\sigma(S_i')}{\bar S_1'}\biggr]^2_\mathrm{SICT}
&\equiv&\frac{1}{S_1^2}\sum_{j=1}^4 
	\left(\epsilon_{ij}\right)_{\rm SICT}^2\,S^2_j \\
&=& \frac{1}{\pi}
	\int_{0}^{\infty} S(\omega)\,
	\bigl(
	\tens{\tilde H}(\omega)\tens{g}\tens{\tilde H}^\dagger(\omega)
	\bigr)_{ii}\;\textrm{d}\omega  
\simeq \frac{1}{\pi} \sum_{j=1}^4 \tens{g}_{jj}
	\int_{0}^{\infty} S(\omega)\,
	\tilde H_{ij}^2(\omega)\;\textrm{d}\omega\;,
\end{eqnarray*}
from which we derive
\begin{equation} \label{eq:errmat_SICT}
\left(\epsilon_{ij}\right)_{\rm SICT}^2
\simeq \frac{1}{\pi} \frac{\tens{g}_{jj}}{(S_j/S_1)^2}
	\int_{0}^{\infty} S(\omega)\,
	\tilde H_{ij}^2(\omega)\;\textrm{d}\omega\;.
\end{equation}
In particular, the last equation shows that the diagonal elements of 
$\bm{\epsilon}$ are essentially due to the spatial-modulation error 
(cf.~Eq.~(\ref{eq:spat_error})), and therefore are practically unaffected by polarization cross-talk.

\section{Conclusions} \label{sec:concl}

Recent technical advances in the nanofabrication of optical metasurfaces
\citep{Ru21} have introduced new ways to perform the polarization
analysis of a signal, which are becoming mature for application to
astrophysical research. The Solar Imaging Metasurface Polarimeter
(SIMPol) is a prototype instrument that was funded by the NASA program
for the Heliophysics Technology and Instrument Development for Science
(H-TIDeS), and implements a metasurface polarization-splitter (MPS) grating
designed to work efficiently in the spectral range around the
\ion{Sr}{1} photospheric line at 460.7\,nm. The design effort aimed at
maximizing the throughput and diattenuation of the MPS into a minimum
number of linearly independent, polarized diffraction orders through
which the full state of polarization of the target can be reconstructed.

These new devices effectively enable the possibility of ``snapshot''
polarimetry for both ground-based and space-borne polarimetric
instrumentation, greatly reducing the risk associated with estimating and
mitigating the effects of image instabilities (atmospheric turbulence,
spacecraft jitter) on the accuracy of the polarization information
acquired by the instrument. They also remove the need for mechanisms
that are instead necessary to perform the temporal modulation of a
polarized signal, such as rotation stages to spin birefringent
optics, and tip-tilt mirrors aimed at stabilizing internally the position
of the image during the modulation cycle, thus eliminating all mission risks associated with the possible failure of such mechanisms.

Because of the growing interest in snapshot polarimetry, with this work 
we wanted to demonstrate how the formalism we developed in \cite{Ca12}, 
to treat the problem of seeing-induced cross-talk in polarization
measurements that adopt a temporal modulation scheme of the signal, can
also be applied to evaluating the impact of image instabilities on spatially
modulating instruments such as SIMPol. A fundamental result we derived
is that the polarimetric errors of such instruments simply correspond to
the loss of spatial and spectral resolutions of the target due to the smearing from
the image motion. On the other hand, no actual polarization cross-talk---i.e., the
measurement bias of a given Stokes parameter $S_i$ when a different
parameter $S_j$ is also non-vanishing---can be produced in such instruments 
during the acquisition of the signal.
Polarization cross-talk can still be introduced as a consequence of an
imperfect co-registration of the independent polarization states prior
demodulation into the Stokes vector of the target, or inadequate flat-fielding of the data. However, this is a 
problem common to \emph{all} polarimeters, including those adopting
temporal modulation schemes, whether dual-beam or single-beam (e.g., 
because of beam wobble caused by a slightly wedged modulator 
optic).\footnote{In dual-beam polarimeters it is sometimes 
possible to implement a \emph{beam swapping} of the signals 
along the modulation cycle, i.e., to channel the same modulated 
signal on both beams at two different modulation steps. In such 
as case one can show that the impact of imperfect flat-fielding is 
significantly reduced \citep{dTI03}. However, this possibility is 
generally limited to very specific polarimeter designs and 
observation targets, the most typical one being a half-wave 
retarder modulator dedicated to the measurement of linear polarization only.}

As polarimetric instruments adopting the more traditional temporal
modulation schemes are not going to disappear any time soon, there is
still an interest in improving our understanding of how image
instabilities impact polarization measurements performed with such
instruments. For this reason, we have provided further insights on the
SICT formalism of \cite{Ca12}, and derived explicit formulas for the
estimation of the mechanical stability tolerances of slit-based
spectro-polarimeters. Finally, we showed how the polarimetric noise 
estimation provided by the SICT formalism can be used to inform 
polarimetric error budgets for the purpose of instrument and mission 
design.

\begin{acknowledgements}
The authors have benefited from discussions with D. M. Harrington (NSO) and N. Rubin (UCSD) on the subjects presented in this paper. This material is based upon work supported by the National Center for Atmospheric Research, which is a major facility sponsored by the National Science Foundation under Cooperative Agreement No.~1852977.
\end{acknowledgements}

\begin{appendix}
\section{Some fundamental properties of the cross-talk integrand}
\label{sec:properties}

We formally demonstrate that the integrand of Eq.~(\ref{eq:variance}) is an even function of $\omega$, so that equation can be recast in the form of Eq.~(\ref{eq:error}). In order to do this, we first observe that
\begin{eqnarray*}
\tens{\tilde H}(\omega)\tens{g}\tens{\tilde H}^\dagger(\omega)
+
\tens{\tilde H}(-\omega)\tens{g}\tens{\tilde H}^\dagger(-\omega)
&=& 
	\tens{\tilde H}(\omega)\tens{g}\tens{\tilde H}^\dagger(\omega)
	+
	\tens{\tilde H}^\ast(\omega)\tens{g}\tens{\tilde H}^t(\omega) \\
&=& \tens{\tilde H}(\omega)\tens{g}\tens{\tilde H}^\dagger(\omega) +
	\bigl(
	\tens{\tilde H}(\omega)\tens{g}\tens{\tilde H}^\dagger(\omega)
	\bigr)^t\;,
\end{eqnarray*}
where in the first row we used the conjugation property of the Fourier transform of a real function, and in the second row we recalled that $\tens{g}$ is symmetric. Because the diagonal elements are invariant under matrix tranposition, the above relation immediately implies
\begin{eqnarray*}
\bigl(
\tens{\tilde H}(\omega)\tens{g}\tens{\tilde H}^\dagger(\omega)
\bigr)_{ii} + \bigl(
\tens{\tilde H}(-\omega)\tens{g}\tens{\tilde H}^\dagger(-\omega)\bigr)_{ii}
&=& 2\bigl(
	\tens{\tilde H}(\omega)\tens{g}\tens{\tilde H}^\dagger(\omega)
	\bigr)_{ii}\;,
\end{eqnarray*}
or
\begin{equation} \label{eq:integrand.alt}
\bigl(
\tens{\tilde H}(-\omega)\tens{g}\tens{\tilde H}^\dagger(-\omega)
\bigr)_{ii}
= \bigl(
	\tens{\tilde H}(\omega)\tens{g}\tens{\tilde H}^\dagger(\omega)
	\bigr)_{ii}\;.
\end{equation}
Together with the condition $S(-\omega)=S(\omega)$, which must be true for any \emph{real} random process, the above identity demonstrates the 
parity of the integrand of Eq.~(\ref{eq:variance}).

Additionally, we want to verify that the diagonal element (\ref{eq:integrand.alt}) is non-negative, as required by the need for the cross-talk variance Eq.~(\ref{eq:variance}) to be a positive quantity for any $S(\omega)$. Such a non-negativity property follows formally and directly 
from the symmetry and non-negativity of $\tens{g}$.
To show this, we first note that
$\tens{g}=\tens{V}\bm{\Lambda}\tens{V}^t$, where $\bm{\Lambda}$ is the 
diagonal matrix of the (non-negative) eigenvalues of $\tens{g}$, and
$\tens{V}$ the (real) orthogonal matrix of its eigenvectors. Therefore,
for any real four-vector $\bm{w}$ we can write,
\begin{eqnarray*}
\bm{w}^t\tens{\tilde H}(\omega)\tens{g}\tens{\tilde H}^\dagger(\omega)\bm{w}
&=&\bm{w}^t\tens{\tilde H}(\omega)\,\tens{V}\bm{\Lambda}	\tens{V}^t\,\tens{\tilde H}^\dagger(\omega)\bm{w} \\
&=&\bigl[\tens{V}^t\tens{\tilde H}^\dagger(\omega)\bm{w}\bigr]^\dagger \bm{\Lambda}\bigl[\tens{V}^t\tens{\tilde H}^\dagger(\omega)\bm{w}\bigr]
\equiv \bm{\tilde{w}}^\dagger\bm{\Lambda}\,\bm{\tilde{w}} \ge 0\;,
\qquad \forall\bm{w}\in\mathbb{R}^4\;,
\end{eqnarray*}
the last inequality being a consequence of the non-negativity of $\tens{g}$.
Because the last expression represents a non-negative diagonal quadratic 
form, each of the diagonal elements (\ref{eq:integrand.alt}) must be non-negative.
This is immediately evident if we set $\bm{w}$ in the last equation to
any of the unit vectors $\bm{e}_i$ of $\mathbb{R}^4$, since 
\begin{equation}
\bigl(
\tens{\tilde H}(\omega)\tens{g}\tens{\tilde H}^\dagger(\omega)
\bigr)_{ii}
=
\bm{e}_i^t\tens{\tilde H}(\omega)\tens{g}\tens{\tilde H}^\dagger(\omega)\bm{e}_i
\ge 0\;,\qquad i=1,2,3,4\;.
\end{equation}

\end{appendix}

\end{document}